\documentclass[aps,prl,superscriptaddress,preprintnumbers,twocolumn,amsmath,amssymb,floatfix]{revtex4-1}
\usepackage{graphicx}
\usepackage{color}
\definecolor{RED}{rgb}{1,0,0}
\definecolor{BLUE}{rgb}{0,0,1}
\begin{document}

\title{Kondo effect and subatomic structures of single U atoms on graphene/6H-SiC(0001)}
\author{W. Feng}
\affiliation{Science and Technology on Surface Physics and Chemistry Laboratory, Mianyang 621908, China}
\affiliation{Institute of Materials, China Academy of Engineering Physics, Mianyang 621908, China}
\author{P. Yang}
\affiliation{Science and Technology on Surface Physics and Chemistry Laboratory, Mianyang 621908, China}
\affiliation{School of Physics, Nankai University, Tianjin 300071, China }

\author{B. K. Yuan}
\affiliation{Key Laboratory of Artificial Structures and Quantum Control, School of Physics and Astronomy, Shanghai Jiaotong University, Shanghai 200240, China}
\author{Z. P. Hu}
\affiliation{School of Physics, Nankai University, Tianjin 300071, China }
\author{X. G. Zhu}
\author{S. Y. Tan}
\author{X. C. Lai}
\affiliation{Science and Technology on Surface Physics and Chemistry Laboratory, Mianyang 621908, China}

\author{Q. Liu}
\email{liuqin493@163.com}
\affiliation{Science and Technology on Surface Physics and Chemistry Laboratory, Mianyang 621908, China}
\affiliation{Institute of Materials, China Academy of Engineering Physics, Mianyang 621908, China}
\author{Q. Y. Chen}
\email{sheqiuyun@126.com}
\affiliation{Science and Technology on Surface Physics and Chemistry Laboratory, Mianyang 621908, China}
\affiliation{Institute of Materials, China Academy of Engineering Physics, Mianyang 621908, China}

\begin{abstract}

The Kondo effect typically arises from the spin-flip scattering between the localized magnetic moment of the impurity and the delocalized electrons in the metallic host, which leads to a variety of intriguing phenomena. Here, by using scanning tunnelling microscopy/spectroscopy (STM/STS), we present the Kondo effect and subatomic features of single U adatom on graphene/6H-SiC(0001). A dip spectral feature can be observed around the Fermi energy, which is termed as the ``fingerprint" of the Kondo resonance in STS;
 in addition, two subatomic features with different symmetries: a three-lobe structure and a donghnut-like structure can be observed from the dI/dV maps.
The Kondo resonance is only detectable within 5~\AA~of the lateral distance from the U atom center, which is much smaller than the distances observed in Co atoms on different surfaces, indicating the more localized 5$f$ states than 3$d$ orbitals. By comparing with density functional theory calculations, we find that the two subatomic features displaying different symmetries originate from the selective hybridization between U 6$d$, 5$f$ orbitals and the $p_z$ orbitals from two inequivalent C atoms of the multilayer graphene.

\end{abstract}

\maketitle

How localized $d$ or $f$ electrons interact with the delocalized conduction electrons is a central question in condensed matter physics, which leads to a variety of intriguing phenomena such as
the anomalous behavior in the resistivity, specific heat, and magnetic susceptibility of dilute magnetic alloys \cite{Coleman.10,Varma.76,Stewart.84}. One of the typical examples is the Kondo effect, which arises from spin-flip scattering between a single magnetic atom and the electrons of a metal host. At sufficiently low temperature, the spin of the surrounding conduction electrons interacts with the spin of the magnetic impurity and causes a correlated screening cloud, so that no net moment remains at the Kondo impurity site. Screening the spin of a magnetic atom via the coupling between the spin and electronic degrees of freedom results in a strongly resonant peak in the density of states around the Fermi energy ($E_F$), the Kondo resonance. The spectral features of the Kondo resonance in $f$-electron based heavy-fermion compounds are often characterized by the large $f$ spectral weight around $E_F$ by angle-resolved photoemission spectroscopy (ARPES) \cite{Qiuyun.Co,Qiuyun.Rh,Qiuyun.Ir}. Scanning tunneling spectroscopy (STS) studies of the single magnetic adatoms on surfaces of nonmagnetic metals observe sharp spectra features around $E_F$ \cite{Jiutao.98,Madhavan.98}, which is normally regarded as the fingerprint of the Kondo effect and interpreted as the quantum interference between two electron-tunneling channels \cite{Gumbsch.10}.

Kondo resonance induced by individual atoms have been successfully observed by STM/STS in many 3$d$ magnetic impurity systems, such as  Ti atoms on Ag(100) \cite{Nagaoka.02}, Co atoms on Au(111)  \cite{Madhavan.98}, Cu(111) and Cu(100) \cite{Knorr.02,Manoharan.00}, Ag(100) and Ag(111) \cite{Wahl.04,Schneider.02}, Cu$_2$N/Cu(100) \cite{Otte.08}, and Ru(0001) \cite{Feng.16}. However, the case of single 4$f$ rare-earth atom is still under debate \cite{Jiutao.98,Ternes.09,Silly.04}. Whereas, research on 5$f$-impurity related systems has never been reported, which is an essential part to facilitate the complete understanding of Kondo physics.

Another important issue is the observation of subatomic features of single atoms adsorbed on different surfaces, which is crucial for understanding the orbital characters and bonding behavior of the system. Although visualizing chemical bond structures in large organic molecules or clusters by atomic force microscopy (AFM) have been widely reported in recent years \cite{Oteyza.13,Zhang.13,Emmrich.15,Jamneala.01}, there are only a few successful cases that directly observe the subatomic features of single atoms \cite{Giessibl.00,Giessibl.01,Herz.03,Emmrich.15,Lian.10}. Substructures of the tip atom was first revealed
by AFM \cite{Giessibl.00}. Later on, similar observations were reported for Si adatoms imaging Si tips \cite{Giessibl.01} and Co$_6$Fe$_3$Sm tips \cite{Herz.03}. Observation of the subatomic structures of the sample atoms instead of the tip atoms was only realized in the following systems: Cu and Fe adatoms on Cu surface \cite{Emmrich.15} by AFM, Pb adatoms on Pb(111) \cite{Lian.10} and Ni adatoms on graphene \cite{Gyamfi.12} by STM. Up to now, distinct observation of the subatomic features in individual surface adatoms is still very challenging, especially for STM.

In the present study, we report the Kondo effect and subatomic features of single U atoms on graphene/6H-SiC(0001) by STM/STS. A dip spectral feature can be observed around $E_F$, which is the fingerprint of the Kondo resonance. The Kondo effect is localized within a radius of 5~\AA~around the U atom center. Meanwhile, subatomic features with a three-lobe structure and a donghnut-like shape are distinctly observed in the dI/dV maps. By comparing with DFT calculations,   we find that these subatomic features are due to the selective hybridization between U atoms and the two inequivalent C atoms of the substrate. Our results present the first distinct observation of the Kondo effect, together with the subatomic features of isolated 5$f$-impurity magnetic atoms adsorbed on nonmagnetic surfaces.


The substrate preparations were performed in an ultrahigh vacuum (UHV) chamber with a base pressure better than $8\times10^{-11}$~mbar. Epitaxial graphene films were grown on 6H-SiC(0001) following a standard procedure \cite{Huang.08}. The depositions of U and all the measurements were conducted in another UHV chamber with a base pressure below $1.5\times10^{-11}$~mbar. A small ingot of 99.9\% purity U metal was held with a molybdenum crucible and heated to degas and reduce impurities for a long time under UHV conditions by e-beam evaporator. Then a small amount of U was deposited onto the graphene/6H-SiC(0001) substrate at 7 K. All the STM and AFM measurements were performed by using a commercial qPlus-equipped STM/AFM at 4.2 K. Clean tungsten tips were used after e-beam heating and being treated on a clean Cu(111) substrate. The dI/dV spectra were collected through a standard lock-in technique by applying a 4 mV modulation with a frequency of 731 Hz to sample bias. AFM images were recorded by detecting the frequency shift of the qPlus resonator in non-contact mode with a tungsten tip.


DFT calculations  were carried out using Vienna Ab initio Simulation Package (VASP)\cite{M01}. Within the projector augmented wave (PAW)\cite{M02} framework, the plane-wave cutoff energy was set to be 400 eV. The exchange-correlation functional was treated within the PBE \cite{M03} version of generalized gradient approximation (GGA).  Our periodic slab model is a 6$\times$6$\times$1 bilayer graphene supercell. To eliminate  interactions between the neighbouring slabs, we set the length of $z$-direction to be 15 $\AA$. The optB86b-vdW \cite{M04} version of van der Waals interaction was included in our calculation, which shortens the distance between two graphene layers. The Brillouin zone was sampled by $\Gamma$ centered 2$\times$2$\times$1 $k$-points mesh, and total energy was well converged to less than 1$\times 10^{-5}$ eV. All the structures were fully relaxed without symmetric constraint until the residual atomic force on each atom was smaller than 0.02 eV/$\AA$. The calculated equilibrium lattice constant of graphene is 2.47 $\AA$ and the distance between two adjacent layers is 3.31 $\AA$, which is close to  experimental values. In order to accurately calculate the electronic structures, spin-orbit coupling (SOC) was also considered in this work. We adopted GGA+$U$ method to deal with the strong Coulomb interaction between the 5$f$ electrons in U atoms. The value of Hubbard $U$ was determined to be 3.27 eV by the linear response approach \cite{Cococcioni.05,Qiu.20}.

\begin{figure}[tbp]
\includegraphics[width=87mm]{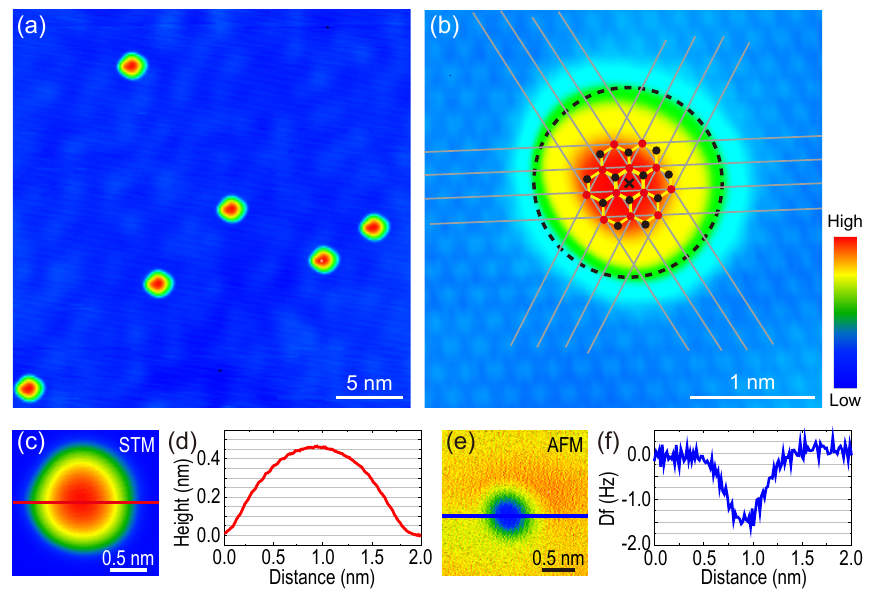}
\caption{ STM topography and AFM image of individual U atoms adsorbed on graphene/6H-SiC(0001). (a) Typical STM constant-current image of individual U atoms  ($V_b$ = -0.6 V, $I$ = 20 pA). (b) Atomically resolved STM image of multilayer graphene surface with a single U adatom on it. Schematic of the lattice structure of the graphene substrate is superimposed on the image of U atom. Red and black dots represent two different sites respectively belong to two inequivalent triangular sublattices on multilayer graphene. Cross and black dashed circle indicate the center and circumference of the U atom, respectively. (c) Zoomed-in STM topographic image of a single U atom ($V_b$ = -0.2 V, $I$ = 10 pA). (d) Height profile measured across the red solid line in panel (c). (e) Non-contact mode AFM frequency shift image of a single U adatom on graphene/6H-SiC(0001) ($V_b$ = -0.2 V, $I$ = 10 pA, tip height $\vartriangle$$z$ = -0.06 nm). (f) Frequency shift $\vartriangle$$f$ as a function of distance measured along the blue solid line in panel (e).
}
\label{FS}
\end{figure}

\begin{figure}
\includegraphics[width=87mm]{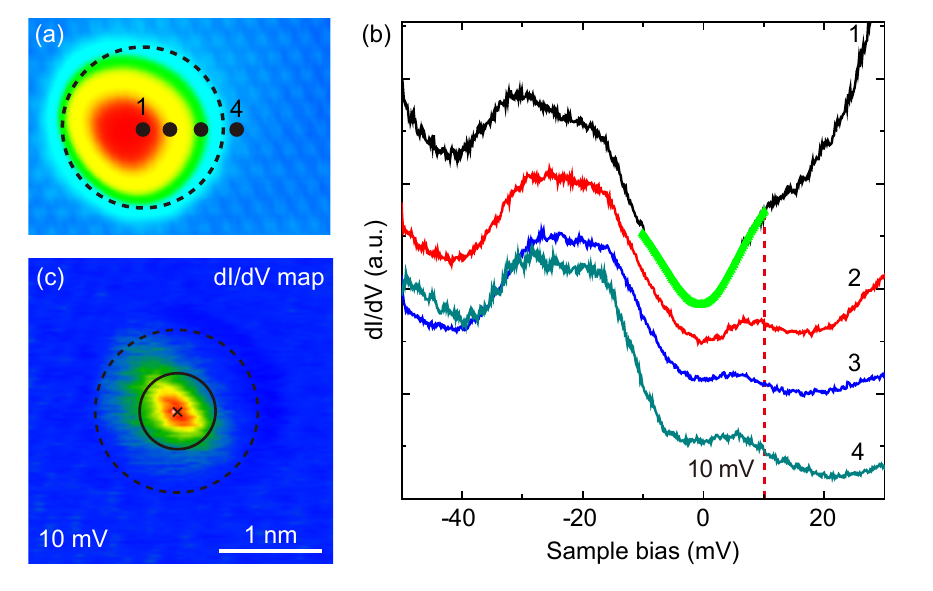}
\caption{Kondo effect of single U adatoms on graphene/6H-SiC(0001). (a) Topographic image of a single U adatom on graphene/6H-SiC(0001). The black dots denote the positions where the dI/dV spectra in panel (b) are taken. (b) dI/dV spectra taken at different positions marked by the black dots in panel (a). Green solid curve is the Fano fit to the asymmetric dip feature around $E_F$. Red dashed line indicates the energy position where the dI/dV map in panel (c) was taken.  (c)  dI/dV map of the single U atom taken at $V_b$= 10 mV and $I$ = 20 pA.}
\label{band}
\end{figure}

Figure 1(a) shows a typical topographic STM image of the epitaxial graphene film grown on graphene/6H-SiC(0001) after the deposition of a small amount of U. Based on close inspection of the substrate surface corrugation in the large scale STM image (see Fig. S1(a) of the supplemental material, SM) and lattice structure in the atomically resolved STM image (Fig. 1(b)), we find the epitaxial graphene thin film has a thickness larger than or at least equal to three layers. As shown in Figs. 1(c) and 1(d), isolated spherical protrusions with a uniform apparent height of 0.45 nm and a diameter of 1.5 nm emerged on graphene/6H-SiC(0001) surface after U deposition. According to literature, a free U atom has an empirical diameter of about 0.31 nm \cite{uranium}, which is comparable with the height of the protrusion but much smaller than its diameter.
To verify the observed spherical protrusion is a single U atom, instead of a cluster of several U atoms, we imaged the protrusion by AFM, which has been proved to be an effective technique for distinguishing cluster from a single atom \cite{Emmrich.15}. Previous studies indicate that the adsorption geometry and inter-atomic chemical bonds would induce specific structures in the AFM image of a cluster, but is usually manifested as a simple single Gaussian peak in STM image. As shown in Figs. 1(e) and 1(f), the spherical protrusion observed in the STM image appears as an intact circular depression without any inner detailed structures in the AFM image and it has a diameter close to 0.5 nm. This further supports the  observed spherical protrusion in STM images only contains a single U atom \cite{Emmrich.15}. Through carefully analyzing the adsorption position of U atom and the atomic lattice structure of substrate, see Fig. 1(b), we find that  U atom adsorbs at the hollow sites of graphene/6H-SiC(0001) surface.

Tunneling spectra taken at different sites (marked by the black dots in Fig. 2(a)) above a single U atom adsorbed on graphene/6H-SiC(0001) are presented in Fig. 2(b).
An asymmetric dip feature locating around  $E_F$ is repeatedly observed in the dI/dV spectra collected around the center of the single U atom.  Fig. 2(c)  displays the dI/dV map of a single U atom taken at $V_b$=10 mV.
Details for the choice of this sample bias can be found in SM.
As shown in Fig. 2(b), this dip feature decreases in amplitude as the STM tip is moved outward from the U atom center, and it is gradually suppressed and completely disappears at about a distance of 5~\AA~from the U atom center, as enclosed by the black solid circles in Fig. 2(c).



The observed dip spectral feature is similar to the Kondo resonance observed by STS on Ce/Ag(111) \cite{Jiutao.98}, Co/Au(111) \cite{Madhavan.98} and Co/Cu(111)\cite{Knorr.02},
and can be reasonably explained as the spectroscopic manifestation of a Kondo resonance state, which can be well described by the so-called Fano equation \cite{Madhavan.01}:
$\rho(\varepsilon)=\frac{\left(q+\varepsilon^{\prime}\right)^{2}}{1+\varepsilon^{\prime 2}}$. Here $\varepsilon^{\prime}=\frac{\varepsilon-\varepsilon_{0}}{\Gamma / 2}$ is normalized energy, $\varepsilon_0$ is energy position of the Kondo resonance relative to $E_F$, $\Gamma$ is full width at half-maximum of resonance curve and $q$ is line-shape parameter \cite{Madhavan.01}. The Fano fits yield a $\Gamma$ = 19.9 meV and $q$ = 0.08, giving a  Kondo temperature $T_K$ of about 114 K. After we subtract the background of graphene, $\Gamma$ reduces to 13 meV, which gives a smaller $T_K$ of 75 K, and these values can be reproduced on different samples, see Figs. S2 and S3. The size of the Kondo cloud of single U atom on graphene/6H-SiC(0001) is much smaller than that of Co atoms on Au(111), Cu(111) and Cu(100) \cite{Madhavan.01, Nagaoka.02}, and the latter reveals a typical size of around 10~\AA. The electronic states relating to the Kondo effect are more localized around the center of U atom than 3$d$-electron system. This is consistent with the more localized nature of 5$f$ orbitals. Fig. S4 shows the spin-polarized partial density of states (PDOS) of the system, from which it is clear that the local magnetic moment of 5.0 $\mu_B$ is mainly from the $f$ orbitals of the U atoms.
Although Kondo effect induced by single 3$d$ transition-metal atoms has been extensively studied before \cite{Madhavan.01, Nagaoka.02,Madhavan.98}, the case of single 4$f$ rare-earth atom is still under debate \cite{Jiutao.98,Ternes.09,Silly.04}, whereas similar studies focusing on single actinide atom with 5$f$ electrons has never been reported before and was first revealed here.

\begin{figure}[tbp]
\includegraphics[width=87mm]{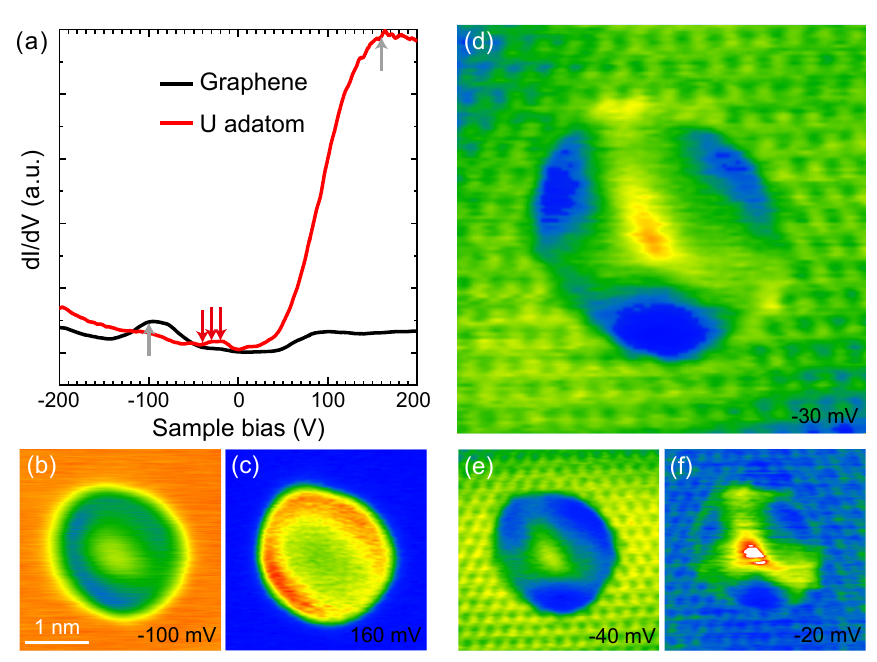}
\caption{Subatomic features of a single U adatom  on graphene/6H-SiC(0001). (a) dI/dV spectra of the adsorbed U atom and the substrate. The gray arrows denote the positions where the dI/dV maps in panels (b) and (c) are taken. The red arrows denote the positions where the dI/dV maps in panels (d-f) are taken.  (b-f) dI/dV maps taken at $I$ = 100 pA, $V_b$ = -100 mV (b); $V_b$ = 160 mV (c); $V_b$ = -30 mV (d); $V_b$ = -40 mV (e); $V_b$ = -20 mV (f).}

\label{resonantPES}
\end{figure}

\begin{figure}
\includegraphics[width=87mm]{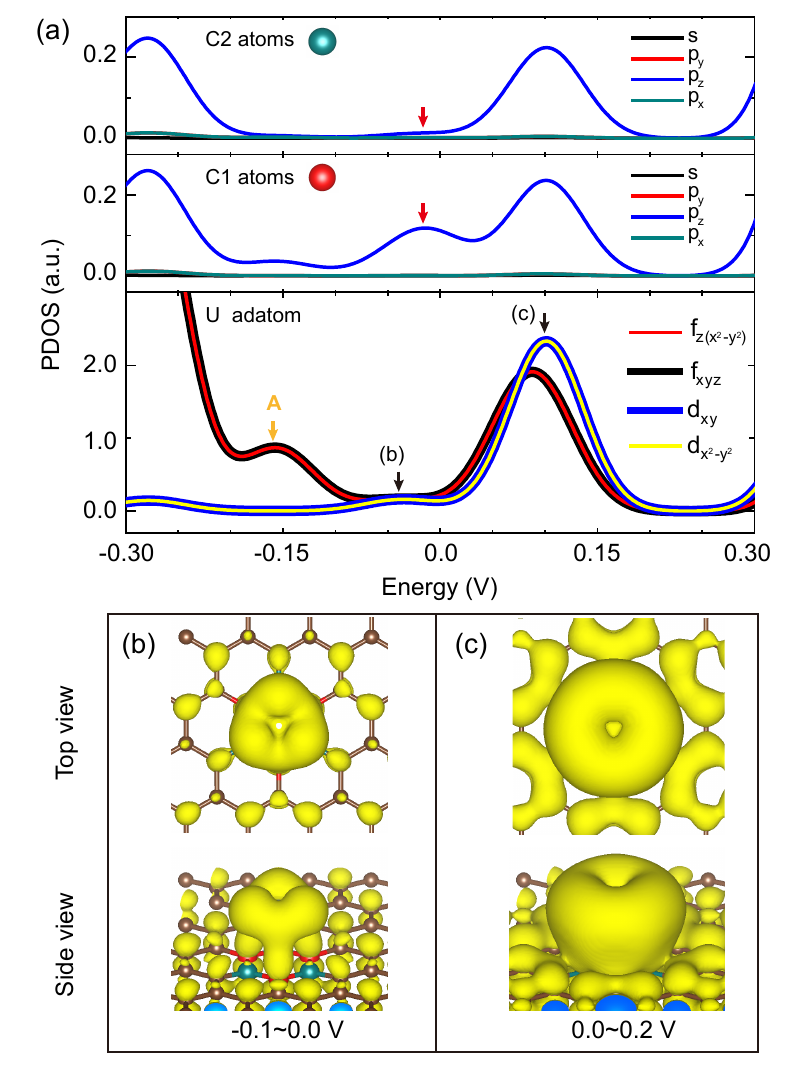}
\caption{PDOS and the isosurfaces of charge density around the U atoms. (a) PDOS of U and C atoms calculated by GGA+$U$+SOC, and red and green atoms represent the two inequivalent C sites. $E_F$ is set to zero. (b) and (c) Isosurfaces of charge density around the U atoms at the energy ranges correspond to the peaks denoted by the arrows in panel (a) with their energy ranges marked below, and the iso-values are  $5.0 \times 10^{-5}$ and $ 1.0 \times 10^{-5} e \AA^{-3}$ for (b) and (c), respectively. }
\label{tdep}
\end{figure}

The line shape of the Fano resonance and the coupling parameter $q$ have been extensively discussed in previous reports \cite{Fano.61,Madhavan.01,Madhavan.98,Jiutao.98}.
If $q$ is large ($|q|\gg1$), the spectrum is Lorentzian and the STM tip is strongly coupled to the atomic orbitals, while
a low value of $q$ ($|q|<1$) implies that the tip is more strongly coupled to the conduction electrons than the atomic orbitals \cite{Madhavan.01}. In the present experiment, the obtained small $q$ value of 0.08 implies that the matrix element for the localized 5$f$ states should be very small, and the dI/dV spectra are more sensitive to the conduction electrons.  The Kondo resonance in STS  reveals itself by reducing the differential conductance for tunneling transmission near $E_F$ and appears as a  Kondo antiresonance in the tunneling spectra.

In addition to the dip spectral feature around $E_F$,
we also observed three hump features in the dI/dV spectra in Fig.~3(a), locating at -100, -30 and 160 mV, respectively. To further investigate the origin of these hump features, dI/dV maps were taken at these energies. From the dI/dV maps taken at -100 and 160 mV in Figs. 3(b) and 3(c), a round and doughnut-like feature can be observed. Figures 3(d-f) present the dI/dV maps taken at three different biases near the hump feature around -30 mV, and they all display three-lobe structure with a threefold symmetry. This three-lobe structure can be best resolved at the sample bias of -30 mV, and the increased intensity at the center of the three lobes in the dI/dV pattern at -20 mV  is due to the influence of the Kondo resonance near $E_F$. The three lobes align in the same direction as the triangular lattice manifested by the STM images of multilayer graphene surface, and the dI/dV patterns exhibit the same symmetry as the substrate. Therefore, we speculate that these special dI/dV patterns reflect the hybridization between the U atoms and the substrate.

To precisely reveal the hybridization mechanism between the U atoms and substrate, we use DFT+$U$ method to calculate the spatial distributions of various atomic orbitals in a single U adatom on graphene. Figure  4 presents the PDOS and the isosurfaces of the charge density around U atom in different energy ranges. From Fig. 4(a), the round feature observed from the dI/dV map at -100 mV in Fig. 3(b) is mainly from the contribution of localized $f$ states as indicated by the orange arrow A in Fig. 4(a), which do not hybridize with the C atoms of the substrate. The three-lobe structure observed in Figs. 3(d-f) is due to the hybridization between the $p_z$ orbitals from one of the two inequivalent C atoms (C1) and the 6$d$ and 5$f$ orbitals of the U atom, as shown in Figs. 4(a) and 4(b). The donghnut-like feature at 160 mV from the dI/dV map in Fig. 3(c) originates from the hybridization between U 6$d$ and 5$f$ orbitals and $p_z$ orbitals of the C atoms from both two sites. In short, both the three-lobe and donghnut-like features originate from the hybridization between the U adatom and C atoms of the substrate. However,  since the donghnut-like structure is due to the hybridization between the U adatom and C atoms from both sides, whereas only one site of the two inequivalent C atoms hybridize with the U adatom in the three-lobe structure, this lowers the symmetry in the three-lobe spectra feature.
Although there are a few reported studies about the observation of subatomic features in a single atom by AFM or STM \cite{Giessibl.00,Giessibl.01,Herz.03,Emmrich.15,Lian.10,Gyamfi.12}, most of them concentrated on the subatomic structures induced by special tip configurations \cite{Giessibl.00,Giessibl.01,Herz.03,Emmrich.15}, while the rest only obtained relatively blurry images or spectral data \cite{Lian.10,Gyamfi.12}. Our STM observation of the clear subatomic orbitals inside a single U adatom is unprecedented. Such a remarkable result does not depend on a special decorated STM tip. Instead, it originates from the large apparent diameter of U adatom and the selective hybridization between U atom and C atoms from two inequivalent sites of the hexagonal lattice on multilayer graphene.
To summarize, our results present the first observation of the Kondo effect, together with the subatomic features of isolated 5$f$-impurtiy U atom on graphene/6H-SiC(0001). The Kondo effect manifests itself by a dip spectral feature near $E_F$, which can be fitted by the Fano equation and the two different subatomic features locating at different energies originate from the selective hybridization between U 5$f$ and 6$d$ orbitals with the $p_z$ orbitals from two inequivalent C atoms of the substrate. 5$f$-electron system takes a special place in the Kondo category, but is less studied. Our results provide new insights into the understanding of the Kondo physics, by extension, the orbital character and bonding behaviors of 5$f$-electron systems.

\begin{acknowledgments}
This work is supported by  the National Science Foundation of China (Grants No.11304291, 11974319, 11874330), the National Key Research and Development Program of China (No. 2017YFA0303104), and the Science Challenge Project (Grants No. TZ2016004).

W. Feng and P. Yang contribute equally.
\end{acknowledgments}

\end{document}